\documentclass[conference]{IEEEtran}
\IEEEoverridecommandlockouts

\usepackage{cite}
\usepackage{amsmath,amssymb,amsfonts}
\usepackage{graphicx}
\usepackage{textcomp}
\usepackage{xcolor}
\usepackage{booktabs}
\usepackage{array}
\usepackage{colortbl}
\graphicspath{{figures/}}
\def\BibTeX{{\rm B\kern-.05em{\sc i\kern-.025em b}\kern-.08em
    T\kern-.1667em\lower.7ex\hbox{E}\kern-.125emX}}
\newcommand{\Dop}{\hat{\mathcal{D}}}
\newcommand{\Nop}{\hat{\mathcal{N}}}
\begin{document}

\title{Time-Domain GNLSE Compact Modeling in SPICE for Electro-Optical Co-Simulation}

\author{%
\IEEEauthorblockN{Pujun Su$^1$, Borong Gu$^2$, Xin Huang$^1$, Danqing Wang$^1$, Guowei Cao$^2$, Yuxuan Li$^3$, Tao Zhang$^4$, and Keren Zhu$^{1,*}$}%
\IEEEauthorblockA{\vspace{2pt}$^1$Fudan University \quad $^2$CUMEC \quad $^3$Shenzhen Empyrean Technology Co., Ltd.\\
$^4$Shanghai Empyrean Information Technology Co., Ltd. \quad $^*$Corresponding author: krzhu@fudan.edu.cn}%
}

\maketitle

\begin{abstract}
As integrated electro-optical systems continue to scale, unified circuit-photonic simulation with high physical fidelity has become essential. Optical envelope evolution in integrated waveguides follows the generalized nonlinear Schr\"{o}dinger equation (GNLSE), which is conventionally integrated via the split-step Fourier method (SSFM) using block-wise fast Fourier transforms (FFTs) over an entire temporal record. Such block-level frequency-domain operations conflict with SPICE-style transient solvers that march sequentially from one time step to the next. To bridge this gap, we introduce a Verilog-A compact waveguide model that directly integrates the GNLSE---capturing distributed higher-order chromatic dispersion coupled with Kerr nonlinearity, two-photon absorption, free-carrier dynamics, and self-steepening---within standard transient circuit analysis. Operating sequentially on incoming time samples under symmetric operator splitting, the formulation replaces FFT-based dispersive propagation with cascaded all-pass filter sections, guaranteeing causality and intrinsic passivity without spectral transforms. Verification in Cadence Spectre confirms agreement with a reference Python implementation within $1.2\times10^{-7}$ of the peak field amplitude. Against Lumerical INTERCONNECT on a 445-nm silicon wire benchmark, the proposed formulation lowers the output intensity error from 63.8\% (linear-only modeling) down to 0.57\%.
\end{abstract}


\section{Introduction}

\begin{figure}[t]
\centering
\includegraphics[width=\columnwidth]{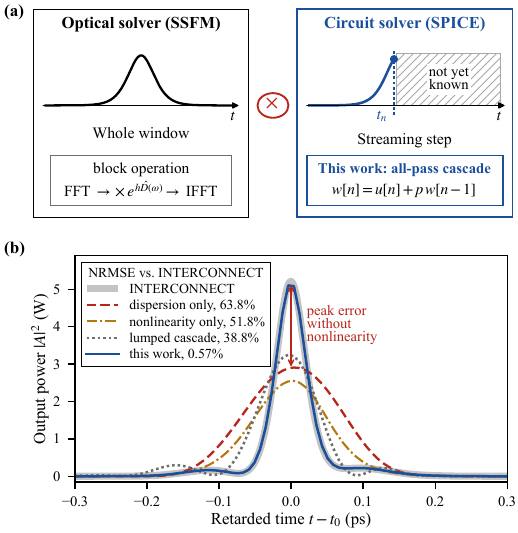}
\caption{Paradigm mismatch between the split-step Fourier method (SSFM) and transient SPICE simulation. (a) The SSFM applies dispersion to a whole time window through FFTs, whereas a transient circuit solver knows the signal only up to the current time point; the proposed model uses a causal all-pass filter (APF) recursion instead. (b) Output power of a 445-nm silicon wire (4.7~mm, 100-fs input, 6~W peak) compared with Lumerical INTERCONNECT.}
\label{fig:motivation}
\end{figure}

Electronic-photonic integrated systems, such as co-packaged optics (CPO)~\cite{minkenbergCopackagedDatacenterOptics2021} and optical I/O (OIO)~\cite{wadeTeraPHYChipletTechnology2020}, are increasingly adopted in data centers and high-performance computing for their low latency, high bandwidth density and energy efficiency~\cite{fangVerilogaBasedTimefrequency2026,chewAdaptiveMicrowavePhotonic2026}.

To verify these heterogeneous systems in one design environment, photonic devices are commonly described by compact models that run natively in electronic design automation (EDA) tools~\cite{soraceagaskarElectroopticalCosimulationIntegrated2015,rhimVerilogABehavioralModel2015,yeTimedomainCompactMacromodeling2019,fangVerilogaBasedTimefrequency2026}. Light propagation in a waveguide, however, is governed by the generalized nonlinear Schr\"{o}dinger equation (GNLSE), which combines frequency-dependent dispersion with power-dependent nonlinearity~\cite{agrawalNonlinearFiberOptics2019,linNonlinearOpticalPhenomena2007}. Its standard solver, the split-step Fourier method (SSFM), alternates between the time and frequency domains over whole time windows~\cite{agrawalNonlinearFiberOptics2019,dudleySupercontinuumGenerationPhotonic2006}, whereas SPICE-class circuit simulators advance point by point in time (Fig.~\ref{fig:motivation}(a)).

\begin{figure*}[t]
\centering
\includegraphics[width=\textwidth]{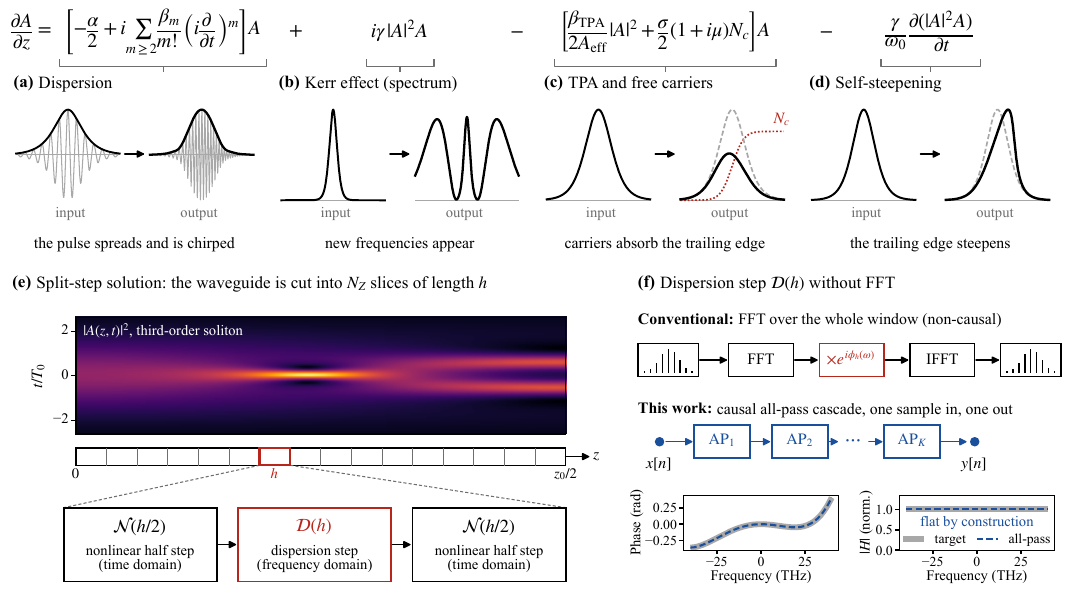}
\caption{The GNLSE term by term and how the proposed model solves it. (a)--(d) Effect of each term on a short pulse, computed with an FFT split-step solver; (c) is a 10-ps, 5-W pulse after 5~mm of silicon, where the dashed curve omits TPA and free carriers and the dotted curve is the normalized carrier density; (d) is exaggerated for visibility. (e) The split-step method cuts the waveguide into $N_Z$ slices, each alternating nonlinear steps in time with a dispersion step in frequency; the map is the computed $|A(z,t)|^2$ of a third-order soliton over half a soliton period $z_0$. (f) The proposed model replaces the FFT dispersion step by $K$ all-pass sections that process one sample at a time; the insets show the fitted phase and the exactly flat magnitude for one step of the 4.7-mm silicon wire ($K=10$).}
\label{fig:gnlse}
\end{figure*}

Existing Verilog-A photonic models therefore capture only one side of the problem. Nonlinear models include the Kerr effect, two-photon absorption (TPA), and free-carrier absorption and dispersion (FCA/FCD), but neglect higher-order dispersion and its distributed interplay with nonlinearity~\cite{zhangVerilogACompactModel2024,zhangCompactPhotonicModel2024}. Frequency-domain macromodels, such as the complex vector fitting of S-parameters by Ye \emph{et al.}~\cite{yeTimedomainCompactMacromodeling2019} and the higher-order dispersion model of Fang \emph{et al.}~\cite{fangVerilogaBasedTimefrequency2026}, are restricted to linear time-invariant (LTI) systems.

The bottleneck is the frequency-domain dispersion step. We observe that vector fitting (VF)~\cite{gustavsenRationalApproximationFrequency1999} can convert a frequency response into a time-domain recursion for transient simulation. Because the dispersion over a short step has a flat magnitude and a phase-only response, we realize it with a cascade of all-pass filters (APFs), so that dispersive and nonlinear steps alternate natively in the time domain.

In this paper, we present the first Verilog-A compact model that solves the GNLSE for distributed, coupled higher-order dispersion and nonlinear propagation inside a SPICE transient analysis. The main contributions of this work are summarized as follows:
\begin{itemize}
    \item \textbf{Coupled Dispersion-Nonlinear Modeling:} A distributed Verilog-A model that captures higher-order dispersion together with Kerr nonlinearity, TPA, free carriers and self-steepening.
    \item \textbf{FFT-Free Time-Domain Dispersion Step via APF Cascade:} Each dispersion step is a low-order all-pass cascade whose magnitude is exact by construction, so long chains stay passive. With 17 to 33 complex MACs per sample and step it matches an exact FFT step, which FIR filters with up to 257 taps and VF models do not.
    \item \textbf{Validated Device- and Circuit-Level Accuracy:} The Spectre implementation matches its Python counterpart to $1.2\times10^{-7}$. Against INTERCONNECT, its waveform error is 0.57\% in a 445-nm silicon waveguide where neglecting nonlinearity yields 63.8\%, and in two complete links it reproduces the design optima that single-effect models miss.
\end{itemize}

\section{Background and Problem Formulation}
An optical pulse in a waveguide is distorted by two kinds of effects: dispersion, which depends on frequency, and nonlinearity, which depends on the instantaneous optical power. The standard solver handles each in the domain where it is simple and alternates between time and frequency over short steps. This section introduces the equation and its solver, and shows why this alternation conflicts with a transient circuit simulator.

\subsection{Generalized Nonlinear Schr\"{o}dinger Equation (GNLSE)}\label{sec:gnlse}
The GNLSE governs how the envelope of an optical pulse evolves along a waveguide~\cite{agrawalNonlinearFiberOptics2019,linNonlinearOpticalPhenomena2007}. It defines the physics the model must reproduce, and its structure, a frequency-domain dispersion operator plus a time-domain nonlinear operator, is what makes it hard to solve inside a circuit simulator.

The equation is written for the slowly varying envelope $A(z,t)$, scaled so that $P=|A|^2$ is the optical power in watts, with $t$ the retarded time:
\begin{equation}
\frac{\partial A}{\partial z} = \Dop A + \Nop(A)\,A, \quad
\Dop = -\frac{\alpha}{2} + i\sum_{m\geq 2}\frac{\beta_m}{m!}\Big(i\frac{\partial}{\partial t}\Big)^{m},
\label{eq:gnlse}
\end{equation}
\begin{equation}
\Nop A = \Big[i\gamma P - \frac{\beta_\mathrm{TPA}}{2A_\mathrm{eff}}P - \frac{\sigma}{2}(1+i\mu)N_c\Big]A - \frac{\gamma}{\omega_0}\frac{\partial (PA)}{\partial t},
\label{eq:nl}
\end{equation}
\begin{equation}
\frac{\partial N_c}{\partial t} = \frac{\beta_\mathrm{TPA}}{2\hbar\omega_0 A_\mathrm{eff}^2}P^2 - \frac{N_c}{\tau_c},
\label{eq:carrier}
\end{equation}
where $\alpha$ is the linear loss, $\beta_m$ the $m$-th order dispersion, $\gamma$ the Kerr parameter, $\beta_\mathrm{TPA}$ the TPA coefficient, $A_\mathrm{eff}$ the effective mode area, $\sigma$ and $\mu$ the FCA cross section and FCD ratio, $N_c$ the free-carrier density, $\tau_c$ the carrier lifetime, and $\omega_0$ the carrier frequency.

Each term produces a characteristic distortion, and the terms interact. Dispersion broadens and chirps the pulse, Kerr self-phase modulation (SPM) broadens its spectrum, TPA clamps the peak while its carriers absorb the trailing edge, and self-steepening skews the peak (Fig.~\ref{fig:gnlse}(a)--(d)). Their interplay is governed by the dispersion length $L_D=T_0^2/|\beta_2|$ and the nonlinear length $L_\mathrm{NL}=1/(\gamma P_0)$ of a pulse with width $T_0$ and peak power $P_0$. When the waveguide length $L$ is comparable to both, dispersion continuously reshapes the SPM-broadened spectrum; a third-order soliton, for example, first compresses to about six times its input peak and then splits (Fig.~\ref{fig:gnlse}(e)).

\begin{figure}[t]
\centering
\includegraphics[width=\columnwidth]{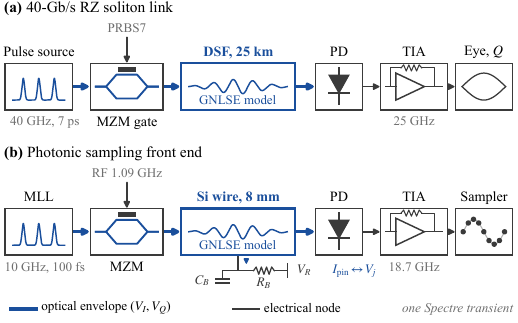}
\caption{Circuit-level testbenches, each simulated in a single Spectre transient. Blue paths carry the optical envelope as $V_I,V_Q$ node voltages and the blue-framed blocks are the proposed GNLSE model; dark paths are electrical nodes. (a) 40-Gb/s RZ soliton link. (b) Photonic sampling front end, with the waveguide's PIN junction tied to the bias network.}
\label{fig:links}
\end{figure}

\subsection{Symmetric Split-Step Fourier Method (SSFM)}\label{sec:ssfm}
The SSFM is the standard numerical solver for the GNLSE~\cite{agrawalNonlinearFiberOptics2019,dudleySupercontinuumGenerationPhotonic2006}. We build on it because it is accurate and already separates the two kinds of physics.

The method applies dispersion and nonlinearity separately over short steps, each in the domain where it is trivial. $\Dop$ is diagonal in frequency and $\Nop$ is local in time, so the SSFM alternates them over steps $h=L/N_Z$; with symmetric Strang splitting~\cite{strangConstructionComparisonDifference1968}, $e^{h(\Dop+\Nop)}=e^{h\Nop/2}e^{h\Dop}e^{h\Nop/2}+\mathcal{O}(h^3)$, and concatenating $N_Z$ steps gives
\begin{equation}
A(L) = \mathcal{N}\big(\tfrac{h}{2}\big)\big[\mathcal{D}(h)\,\mathcal{N}(h)\big]^{N_Z-1}\mathcal{D}(h)\,\mathcal{N}\big(\tfrac{h}{2}\big)\,A(0)
\label{eq:chain}
\end{equation}
with global accuracy $\mathcal{O}(h^2)$. The dispersion step $\mathcal{D}(h)$ is normally an FFT, a multiplication by
\begin{equation}
H_h(\omega) = \exp\Big[h\Big(-\frac{\alpha}{2}+i\sum_{m\geq2}\frac{\beta_m}{m!}\omega^m\Big)\Big] = g\,e^{i\phi_h(\omega)},
\label{eq:Hh}
\end{equation}
with $g=e^{-\alpha h/2}$, and an inverse FFT, which needs the entire time window at once.

\subsection{Causality Constraint and Limitations of Prior Models}\label{sec:gap}
A transient circuit simulator can only use signal values up to the current time point. This causality rule is why the FFT-based SSFM cannot run inside it, and why prior compact models give up either dispersion or nonlinearity.

The FFT dispersion step violates this rule. A Verilog-A module evaluated at $t_n$ sees its inputs only up to $t_n$, and Newton iterations may revisit $t_n$ before the step is accepted. A dispersion step that needs the whole window, as in \eqref{eq:Hh}, is therefore not causal.

Prior compact models avoid the conflict by keeping only one kind of physics, and thereby lose the coupling between the two. Linear macromodels describe the whole waveguide as an LTI filter, through inverse-DFT FIR filters~\cite{fangVerilogaBasedTimefrequency2026} or rational fits~\cite{yeTimedomainCompactMacromodeling2019}, and omit every power-dependent term; nonlinear models evaluate Kerr, TPA and free carriers in time~\cite{zhangVerilogACompactModel2024,zhangCompactPhotonicModel2024} and omit dispersion. Cascading one block of each~\cite{soraceagaskarElectroopticalCosimulationIntegrated2015} replaces the exact propagator by a single split step over the whole length. By the Baker--Campbell--Hausdorff formula,
\begin{equation}
e^{L(\Dop+\Nop)} = e^{L\Dop}\,e^{L\Nop}\,e^{-\frac{L^2}{2}[\Dop,\Nop]+\mathcal{O}(L^3)},
\label{eq:bch}
\end{equation}
so the cascade error is governed by the commutator $[\Dop,\Nop]$, which is nonzero whenever both effects are present. Its relative size is about $L^2/(2L_DL_\mathrm{NL})$, of order one exactly where the effects interact: for the 4.7-mm silicon wire of Section~\ref{sec:results} ($L_D=4.6$~mm) it grows from 0.30 to 1.83 between 1 and 6~W. Splitting into $N_Z$ steps reduces the error to $\mathcal{O}(L^3/N_Z^2)$, but only if the dispersion step stays accurate, cheap and passive when applied hundreds of times: a repeated FIR block multiplies taps, storage and latency by $N_Z$ and accumulates its truncation error, and a rational fit does not bound its gain.

Causal dispersion filters exist only outside circuit simulation: digital back-propagation runs the SSFM in receiver DSP~\cite{ipCompensationDispersionNonlinear2008,fougstedtTimedomainDigitalBack2017}, and IIR or all-pass filters compensate chromatic dispersion digitally~\cite{goldfarbChromaticDispersionCompensation2007} and optically~\cite{madsenOpticalAllpassFilters1998}. The enabling observation is visible in \eqref{eq:Hh}: $|H_h(\omega)|=g$ at all frequencies, so each dispersion step is a pure phase filter that a causal all-pass cascade can realize.

\section{Proposed Causal All-Pass Split-Step Model}
The model keeps the SSFM chain \eqref{eq:chain} but changes how the signal moves through it (Fig.~\ref{fig:gnlse}(e), (f)). The waveguide is cut into $N_Z$ slices, each applying a nonlinear half step, a dispersion step and another nonlinear half step. Instead of collecting a time window and transforming it with FFTs, the model pushes one sample through all slices at every time point of the circuit simulator. Each dispersion step becomes a small all-pass digital filter that uses only past samples and can never amplify the signal; each nonlinear step becomes a local update over a few neighboring samples. Because every operation is causal, the optical solver runs as an ordinary Verilog-A module inside a Spectre transient, next to the electrical circuits. The filter coefficients are fitted once, offline, for each waveguide.

\subsection{Streaming Split-Step Architecture in Verilog-A}\label{sec:stream}
The model is a Verilog-A module that processes the optical signal as a stream of samples rather than as a block. This is necessary because the circuit simulator exchanges data with the module only through node voltages, at time points it chooses itself.

The envelope travels as two node voltages, and each time point carries one sample through all slices. The voltages carry $\mathrm{Re}\,A$ and $\mathrm{Im}\,A$ in $\sqrt{\mathrm{W}}$ and connect directly to modulators and photodetectors. The module holds all $N_Z$ stages of \eqref{eq:chain}, each with its own $K$ all-pass states, a nine-sample ring buffer and a local carrier density $N_{c,s}$. At every sampling instant $t_n=t_0+n\Delta t$, one envelope sample traverses all stages in turn, alternating an all-pass dispersion step and a nonlinear step.

\begin{figure*}[t]
\centering
\includegraphics[width=\textwidth]{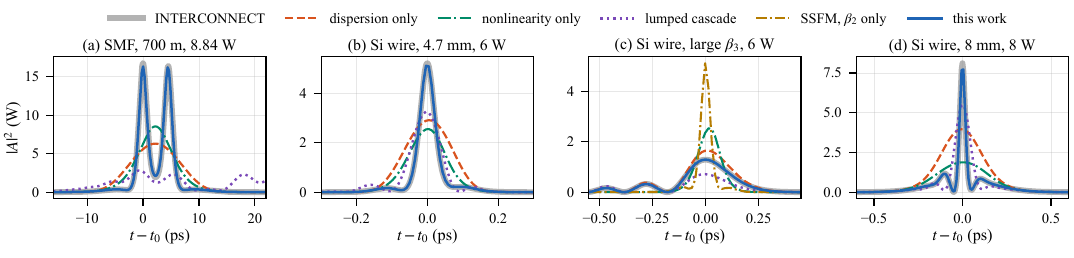}
\caption{Output power waveforms at the highest launch power of each scenario in Table~\ref{tab:main}. The thick gray curve denotes Lumerical INTERCONNECT; the $\beta_2$-only SSFM baseline is shown in (c), where third-order dispersion ($\beta_3$) dominates.}
\label{fig:device}
\end{figure*}

\begin{table*}[t]
\caption{Main Result: Output Power Waveform NRMSE (\%) Against Lumerical INTERCONNECT for All Scenarios and Launch Powers}
\label{tab:main}
\centering
\setlength{\tabcolsep}{3.0pt}
\begin{tabular}{@{}l*{12}{>{\centering\arraybackslash}p{1.08cm}}@{}}
\toprule
 & \multicolumn{3}{c}{SMF, 700 m, 7 ps} & \multicolumn{3}{c}{Si wire, 4.7 mm, 100 fs} & \multicolumn{3}{c}{Si wire, 4.7 mm, large $\beta_3$} & \multicolumn{3}{c}{Si wire, 8 mm, 250 fs} \\
 & \multicolumn{3}{c}{\footnotesize $\Delta t$=244 fs, $N_Z$=800, $K$=8} & \multicolumn{3}{c}{\footnotesize $\Delta t$=7.8 fs, $N_Z$=200, $K$=10} & \multicolumn{3}{c}{\footnotesize $\Delta t$=15.6 fs, $N_Z$=200, $K$=12} & \multicolumn{3}{c}{\footnotesize $\Delta t$=9.8 fs, $N_Z$=200, $K$=16} \\
\cmidrule(lr){2-4}\cmidrule(lr){5-7}\cmidrule(lr){8-10}\cmidrule(l){11-13}
Peak power $P_0$ (W) & 0.98 & 3.93 & 8.84 & 1 & 3 & 6 & 1 & 3 & 6 & 2 & 5 & 8 \\
\midrule
Dispersion only~\cite{fangVerilogaBasedTimefrequency2026} & 22.31 & 77.96 & 65.11 & 10.07 & 32.32 & 63.84 & 4.07 & 12.84 & 28.05 & 24.66 & 65.86 & 96.21 \\
Nonlinearity only & 0.38 & 69.73 & 63.01 & 14.56 & 16.21 & 51.83 & 84.31 & 75.30 & 67.87 & 21.25 & 58.18 & 76.06 \\
Lumped cascade & 7.37 & 70.07 & 86.25 & 4.41 & 12.05 & 38.79 & 3.70 & 19.00 & 46.92 & 11.50 & 33.31 & 62.70 \\
SSFM, $\beta_2$ only & 0.07 & 1.24 & 4.21 & 1.99 & 6.11 & 6.26 & 68.44 & 92.62 & 137.33 & 2.37 & 9.25 & 8.28 \\
\midrule
\rowcolor{black!7}\textbf{This work} & \textbf{0.03} & \textbf{0.35} & \textbf{2.79} & \textbf{0.06} & \textbf{0.23} & \textbf{0.57} & \textbf{0.13} & \textbf{0.18} & \textbf{0.30} & \textbf{0.17} & \textbf{0.70} & \textbf{1.20} \\
\bottomrule
\multicolumn{13}{@{}p{0.98\textwidth}@{}}{\footnotesize $\mathrm{sech}$ inputs. SMF: $D=16$~ps/(nm$\cdot$km), $\beta_{2,3}$ and Kerr, 1, 4 and 9 times the soliton power. Si wires ($445\times220$~nm, 1537~nm): fitted $\beta(\omega)$, loss, Kerr, self-steepening, TPA and FCA/FCD; large $\beta_3=-0.73$~ps$^3$/m. $M=K$.}
\end{tabular}
\end{table*}

\subsection{Passive All-Pass Dispersion Step}\label{sec:apf}
Each dispersion step is realized by a cascade of $K$ first-order all-pass filters. All-pass filters are causal and have the same gain at every frequency, which matches the physics: a dispersion step only delays different frequencies by different amounts and leaves their power unchanged. Only the phase therefore needs to be fitted, and a chain of hundreds of steps can never amplify the signal.

The filter is a product of first-order sections whose magnitude is fixed by construction. With $\theta=\omega\Delta t$, each step must realize $H_h(\theta/\Delta t)$ up to a bulk delay of $M$ samples. We use
\begin{equation}
H_\mathrm{ap}(z) = g\,e^{ic}\prod_{k=1}^{K}\frac{z^{-1}-p_k^{*}}{1-p_k z^{-1}}, \qquad |p_k|<1,
\label{eq:ap}
\end{equation}
with constant phase $c$. Every section has unit modulus on the unit circle, so $|H_\mathrm{ap}(e^{i\theta})|=g$ at all frequencies regardless of fitting accuracy, and a chain of $N_Z$ steps is passive. Each section contributes $2\pi$ of integrated group delay, and $M=K$ keeps the in-band target delay positive.

The poles are fitted offline to the target phase over the signal band: $\{p_k\}$ and $c$ minimize
\begin{equation}
\sum_{\theta\in\Theta_b}\Big|H_\mathrm{ap}(e^{i\theta}) - H_h\big(\tfrac{\theta}{\Delta t}\big)e^{-i\theta M}\Big|^2
\label{eq:fit}
\end{equation}
over samples $\Theta_b$ of the signal band $f_b$, solved by Levenberg--Marquardt with $|p_k|\le 0.98$.

At run time, each section costs one complex state and two multiply-accumulates (MACs) per sample:
\begin{equation}
w_k[n] = u_k[n] + p_k w_k[n-1], \quad y_k[n] = w_k[n-1] - p_k^{*}w_k[n],
\label{eq:aprec}
\end{equation}
with $u_{k+1}=y_k$, so a stage costs $2K+1$ complex MACs and $K$ states, versus $N_\mathrm{tap}$ of each for an FIR filter.

\subsection{Causal Exponential-Midpoint Nonlinear Step}\label{sec:nlstep}
The nonlinear step updates each sample with the power-dependent effects: Kerr phase, TPA, free carriers and self-steepening. These effects are local in time and thus suit a streaming solver; the only difficulty is self-steepening, whose time derivative needs a sample that has not arrived yet.

The step is integrated by a second-order exponential midpoint rule that is exact for pure Kerr SPM. Writing $\Nop A=g(A)A+q(A)$ with $g(A)=i\gamma P-\frac{\beta_\mathrm{TPA}}{2A_\mathrm{eff}}P-\frac{\sigma}{2}(1+i\mu)N_c$ and $q(A)=-\frac{\gamma}{\omega_0}\partial_t(PA)$,
\begin{equation}
\begin{aligned}
A_m &= A\,e^{hg(A)/2} + \tfrac{h}{2}q(A),\\
A^{+} &= \big(A\,e^{hg(A_m)/2} + h\,q(A_m)\big)\,e^{hg(A_m)/2}.
\end{aligned}
\label{eq:expmid}
\end{equation}

Causality is restored by a short, fixed delay. The self-steepening derivative uses a fourth-order central difference, so each stage evaluates $A_m$ at $n-2$ and $A^+$ at $n-4$. The carrier density follows $N_c[n]=N_c[n-1]e^{-\Delta t/\tau_\mathrm{eff}}+G\,P[n-4]^2\Delta t$ with $G=\beta_\mathrm{TPA}/(2\hbar\omega_0A_\mathrm{eff}^2)$. The total latency $D_\mathrm{alg}=[N_Z M + 4(N_Z+1)]\Delta t$ is a known constant; an output delay line removes it or replaces it by the transit time $n_gL/c$. Since all stage states lie within $D_\mathrm{alg}$ of each other, far below carrier lifetimes and bias time constants, they refer to the same circuit time.

\subsection{Event-Driven SPICE Integration and Electro-Optic Coupling}\label{sec:coupling}
The module must run correctly under the simulator's own time stepping and exchange signals with electrical circuits. Otherwise, Newton iterations would advance the filter states repeatedly, and the optics could not respond to the circuit bias.

All discrete-time updates run exactly once per sample. They sit inside an \texttt{@(timer)} event of period $\Delta t$, which forces a time point at every $t_n$ and fires once per scheduled instant rather than per Newton iteration (verified in Section~\ref{sec:setup}). Between events SPICE integrates the circuit freely while optical outputs are held, and idle stages are skipped while their carriers decay analytically.

The circuit couples back to the optics through the carrier lifetime and the junction current. With a PIN junction, the junction voltage $V_j$ sets $\tau_\mathrm{eff}^{-1}=\tau_c^{-1}+k_\mathrm{sw}(V_j+V_\mathrm{bi})$~\cite{turnerfosterUltrashortFreecarrierLifetime2010}, and the swept-out carriers form a current $I_\mathrm{pin}=q_e A_\mathrm{eff}k_\mathrm{sw}(V_j+V_\mathrm{bi})\int N_c\,dz$ injected into the bias network (Fig.~\ref{fig:links}(b)).

\begin{figure}[t]
\centering
\includegraphics[width=0.86\columnwidth]{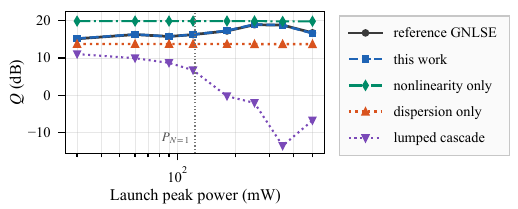}
\caption{40-Gb/s RZ soliton link over 25~km of dispersion-shifted fiber: receiver $Q$ factor versus launch peak power ($P_{N=1}=123$~mW is the fundamental soliton power). The proposed model reproduces the optimum near 250~mW ($Q=19.0$~dB vs.\ reference $19.1$~dB).}
\label{fig:caseb}
\end{figure}

\begin{figure*}[t]
\centering
\includegraphics[width=0.94\textwidth]{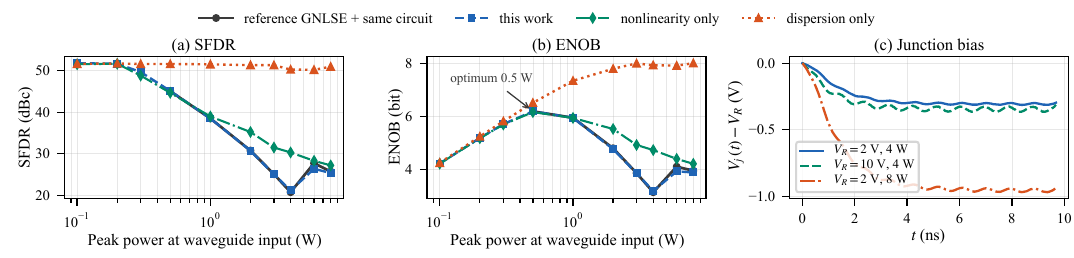}
\caption{Photonic sampling front end at 10~GS/s. (a) SFDR and (b) ENOB (including receiver thermal noise) versus peak power at the waveguide input ($V_R=2$~V). (c) Dynamic PIN junction bias droop $V_j(t)-V_R$ predicted by the proposed coupled model.}
\label{fig:casea}
\end{figure*}

\begin{figure*}[!t]
\centering
\includegraphics[width=0.94\textwidth]{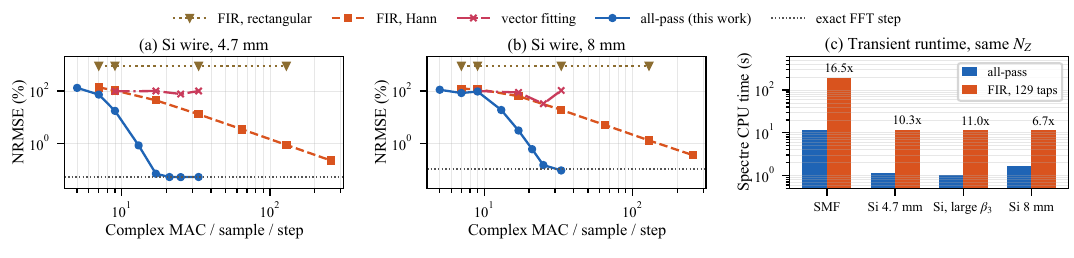}
\caption{Dispersion-step ablation. (a), (b) Waveform NRMSE against a converged FFT reference versus complex MACs per sample and step at $N_Z=200$; dotted line: same model with an exact FFT dispersion step. (c) Spectre CPU time at the highest power of each scenario, all-pass cascade versus 129-tap Hann FIR at equal $N_Z$.}
\label{fig:ablation}
\end{figure*}

\subsection{Discretization and Filter Order Selection}\label{sec:params}
Three parameters set accuracy and cost: the sampling interval $\Delta t$, the step count $N_Z$ and the filter order $K$. Each controls a separate error source, so they can be chosen one after another without a joint search.

Each parameter is raised until its own error falls below the target. We choose $\Delta t$ so that the broadened spectrum lies inside $f_b<1/(2\Delta t)$, raise $N_Z$ until the splitting error meets the target, and raise $K$ until the all-pass error falls below it. An overly fine $\Delta t$ is counterproductive, as the group delay in samples, and hence $K$, grows: in the high-$\beta_3$ wire, $\Delta t=7.8$~fs with $K=16$ leaves 24.6\% error, while $\Delta t=15.6$~fs reaches 0.29\% with $K=12$.

\section{Experimental Results}\label{sec:results}
The experiments answer three questions. Is the model accurate against a commercial photonic simulator (Section~\ref{sec:device}, Table~\ref{tab:main})? Does this accuracy lead to correct design decisions when the model runs inside complete electronic-photonic links (Section~\ref{sec:circuit})? And is the all-pass dispersion step better than the FIR and VF realizations used in prior macromodels (Section~\ref{sec:ablation})?

\subsection{Experimental Setup and Baselines}\label{sec:setup}
We compare the model with a commercial simulator, an independent numerical solver, and four baselines that each mimic a class of prior models by omitting part of the physics. All baselines share our implementation, so every difference stems from the physics.

All transients run in Cadence Spectre 23.1 on an Intel Xeon Gold 6230N node; runtimes are transient CPU times. For the silicon wire, $\beta(\omega)$, $A_\mathrm{eff}=0.194~\mu$m$^2$ and $\gamma=127$~W$^{-1}$m$^{-1}$ come from an eigenmode sweep, with TPA and carrier parameters from~\cite{yinImpactTwophotonAbsorption2007}.

Accuracy is measured as a normalized waveform error against two references. We use $\mathrm{NRMSE} = \|P_\mathrm{model}-P_\mathrm{ref}\|_2/\|P_\mathrm{ref}\|_2$ after cross-correlation alignment, against INTERCONNECT or a converged FFT split-step solver (RK4, spectral self-steepening, 2--8$\times$ finer grid). Across all runs, the Verilog-A envelope matches a Python model of the same discrete algorithm to $1.2\times10^{-7}$ of the peak, which verifies the implementation. Baselines: \emph{dispersion only}, one full-length FIR filter~\cite{fangVerilogaBasedTimefrequency2026}; \emph{nonlinearity only}, which replaces each dispersion step by its loss $g$; \emph{lumped cascade}, one full-length nonlinear and one dispersive block; and \emph{SSFM, $\beta_2$ only}, which omits $\beta_3$ and higher orders.

\subsection{Device-Level Waveform Accuracy}\label{sec:device}
We first compare output waveforms of single waveguides against INTERCONNECT. This is the core test, since every circuit-level result rests on it.

The proposed model is the most accurate at all twelve power levels (Table~\ref{tab:main}). It stays at or below 1.20\% in every silicon case. Its largest deviation, 2.79\% for the third-order fiber soliton, is shared by the converged solver, which differs from INTERCONNECT by 2.86\% and from our model by 0.09\%. Each run takes 1 to 11~s.

Each baseline fails where its omitted physics dominates (Fig.~\ref{fig:device}). Without nonlinearity the pulse broadens instead of compressing (Fig.~\ref{fig:motivation}(b)); without dispersion only the fundamental soliton is reproduced; the lumped cascade creates side pulses, its error rising with power as predicted by \eqref{eq:bch}; and omitting $\beta_3$ yields a symmetric 5.1-W peak instead of a 1.3-W asymmetric pulse.

\subsection{Circuit-Level Electronic-Photonic Co-Simulation}\label{sec:circuit}
We then simulate the two complete links of Fig.~\ref{fig:links}, each in one Spectre transient with its modulator, photodiode, transimpedance amplifier (TIA) and sampler. This tests whether waveform accuracy carries over to the metrics a circuit designer uses: SFDR, ENOB and $Q$.

The reference isolates the optical model from the receiver. As INTERCONNECT cannot simulate circuits, the converged solver's optical output is replayed into the identical receiver netlist, so metric differences stem only from the optics. SFDR and ENOB use 64 coherent samples after 32 warm-up pulses with receiver noise; $Q$ is taken at the widest-eye phase.

\emph{1) 40-Gb/s RZ soliton link:} the model reproduces the optimal launch power, where all baselines fail qualitatively. The 25-km dispersion-shifted fiber spans $4L_D$ ($\Delta t=0.39$~ps, $N_Z=800$, $K=6$). The model tracks the reference $Q$ within 0.15~dB and reproduces the optimum near $2P_{N=1}$, which offsets the 5-dB fiber loss (Fig.~\ref{fig:caseb}). The 160-bit transient runs in 26~s.

\emph{2) Photonic sampling front end:} the model finds the power optimum and the bias droop that single-effect models miss. The 8-mm silicon wire carries a PIN junction biased at $V_R=2$~V through $R_B=500~\Omega$ and $C_B=2$~pF. The model tracks the reference within 0.5~dB in SFDR and 0.06~bit in ENOB up to 4~W and finds the optimum at 0.5~W, while the linear model predicts ENOB rising monotonically and the nonlinearity-only model overestimates SFDR by up to 9.6~dB (Fig.~\ref{fig:casea}). At 8~W the swept-out carriers pull the junction from 2 to 1.03~V within 2~ns, which lengthens the carrier lifetime from 0.38 to 0.54~ns; a model driven by the nominal bias would underestimate it by 30\%. A run takes 224~s (620~s without stage skipping).

\subsection{Dispersion Filter Ablation and Runtime}\label{sec:ablation}
Finally, we isolate the dispersion step, the component that makes the solver causal. We replace it in the same model by the windowed FIR of~\cite{fangVerilogaBasedTimefrequency2026} and by a passivity-enforced VF model~\cite{gustavsenRationalApproximationFrequency1999}, with the same band, weighting and bulk delay (residues refitted under $|H|\le g$).

The all-pass cascade reaches the accuracy of an exact FFT step at the lowest cost. It does so with 17 to 33 MACs (Fig.~\ref{fig:ablation}(a), (b)). The Hann FIR stays above this floor even with 257 taps and saturates (0.90\% and 1.3\% for 129 taps at any $N_Z\ge100$) as its truncation error accumulates; the rectangular FIR diverges; VF stays at 30--100\% because its pole-residue form does not tie the magnitude to $g$.

The all-pass step is also an order of magnitude faster. At equal $N_Z$ it runs $6.7\times$ to $16.5\times$ faster than a 129-tap FIR (Fig.~\ref{fig:ablation}(c)), and a 257-tap FIR remains less accurate in the high-$\beta_3$ wire (1.8\% vs.\ 0.13--0.30\%) while being $42\times$ to $49\times$ slower. The poles are fitted once per waveguide and step size (about 80~s for $K=10$).
\section{Conclusion}
We presented a Verilog-A model that solves the GNLSE inside SPICE transient analysis with causal, passive all-pass dispersion steps. It stays within 1.2\% of INTERCONNECT where linear models reach 96\%, and predicts link optima and bias droop that single-effect models miss. Its scope is unidirectional, single-band propagation within $f_b$.

\newpage
\bibliographystyle{IEEEtran}
\bibliography{references}

@article{chewAdaptiveMicrowavePhotonic2026,
  title = {Adaptive Microwave Photonic Interference Mitigation Technique for Dynamic {{RF}} Environment},
  author = {Chew, Suen Xin and Daniel, Nicole and Chen, Hongchao and Nguyen, Linh and Yi, Xiaoke},
  year = 2026,
  month = jan,
  journal = {Journal of Lightwave Technology},
  volume = {44},
  number = {2},
  pages = {447--457},
  issn = {0733-8724, 1558-2213},
  doi = {10.1109/JLT.2025.3632273},
  urldate = {2026-03-03},
  copyright = {https://creativecommons.org/licenses/by-nc-nd/4.0/},
  langid = {english},
}

@article{fangVerilogaBasedTimefrequency2026,
  title = {A {Verilog-A} Based Time-Frequency Domain Modeling Method for Dispersive Photonic Devices in Electronic-Photonic Convergence},
  author = {Fang, Hao and Liu, Qiang and Feng, Jiajun and Xu, Xiao and Du, Liuge and Zhao, Jia},
  year = 2026,
  month = mar,
  journal = {Journal of Lightwave Technology},
  volume = {44},
  number = {6},
  pages = {2436--2448},
  issn = {0733-8724, 1558-2213},
  doi = {10.1109/JLT.2025.3642856},
  urldate = {2026-09-03},
  copyright = {https://ieeexplore.ieee.org/Xplorehelp/downloads/license-information/IEEE.html},
  langid = {english},
}

@book{agrawalNonlinearFiberOptics2019,
  title = {Nonlinear Fiber Optics},
  author = {Agrawal, Govind P.},
  edition = {6},
  year = {2019},
  publisher = {Academic Press},
  address = {London}
}

@article{dudleySupercontinuumGenerationPhotonic2006,
  title = {Supercontinuum Generation in Photonic Crystal Fiber},
  author = {Dudley, John M. and Genty, Go{\"e}ry and Coen, St{\'e}phane},
  journal = {Reviews of Modern Physics},
  volume = {78},
  number = {4},
  pages = {1135--1184},
  year = {2006},
  doi = {10.1103/RevModPhys.78.1135}
}

@article{linNonlinearOpticalPhenomena2007,
  title = {Nonlinear Optical Phenomena in Silicon Waveguides: Modeling and Applications},
  author = {Lin, Q. and Painter, Oskar J. and Agrawal, Govind P.},
  journal = {Optics Express},
  volume = {15},
  number = {25},
  pages = {16604--16644},
  year = {2007},
  doi = {10.1364/OE.15.016604}
}

@article{yinImpactTwophotonAbsorption2007,
  title = {Impact of Two-Photon Absorption on Self-Phase Modulation in Silicon Waveguides},
  author = {Yin, Lianghong and Agrawal, Govind P.},
  journal = {Optics Letters},
  volume = {32},
  number = {14},
  pages = {2031--2033},
  year = {2007},
  doi = {10.1364/OL.32.002031}
}

@article{strangConstructionComparisonDifference1968,
  title = {On the Construction and Comparison of Difference Schemes},
  author = {Strang, Gilbert},
  journal = {SIAM Journal on Numerical Analysis},
  volume = {5},
  number = {3},
  pages = {506--517},
  year = {1968},
  doi = {10.1137/0705041}
}

@article{gustavsenRationalApproximationFrequency1999,
  title = {Rational Approximation of Frequency Domain Responses by Vector Fitting},
  author = {Gustavsen, Bjorn and Semlyen, Adam},
  journal = {IEEE Transactions on Power Delivery},
  volume = {14},
  number = {3},
  pages = {1052--1061},
  year = {1999},
  doi = {10.1109/61.772353}
}

@article{soraceagaskarElectroopticalCosimulationIntegrated2015,
  title = {Electro-Optical Co-Simulation for Integrated {CMOS} Photonic Circuits with {VerilogA}},
  author = {Sorace-Agaskar, Cheryl and Leu, Jonathan and Watts, Michael R. and Stojanovic, Vladimir},
  journal = {Optics Express},
  volume = {23},
  number = {21},
  pages = {27180--27203},
  year = {2015},
  doi = {10.1364/OE.23.027180}
}

@article{turnerfosterUltrashortFreecarrierLifetime2010,
  title = {Ultrashort Free-Carrier Lifetime in Low-Loss Silicon Nanowaveguides},
  author = {Turner-Foster, Amy C. and Foster, Mark A. and Levy, Jacob S. and Poitras, Carl B. and Salem, Reza and Gaeta, Alexander L. and Lipson, Michal},
  journal = {Optics Express},
  volume = {18},
  number = {4},
  pages = {3582--3591},
  year = {2010},
  doi = {10.1364/OE.18.003582}
}

@article{yeTimedomainCompactMacromodeling2019,
  title = {Time-Domain Compact Macromodeling of Linear Photonic Circuits via Complex Vector Fitting},
  author = {Ye, Yinghao and Spina, Domenico and Deschrijver, Dirk and Bogaerts, Wim and Dhaene, Tom},
  journal = {Photonics Research},
  volume = {7},
  number = {7},
  pages = {771--782},
  year = {2019},
  doi = {10.1364/PRJ.7.000771}
}

@article{wadeTeraPHYChipletTechnology2020,
  title = {{TeraPHY}: A Chiplet Technology for Low-Power, High-Bandwidth In-Package Optical {I/O}},
  author = {Wade, Mark T. and others},
  journal = {IEEE Micro},
  volume = {40},
  number = {2},
  pages = {63--71},
  year = {2020},
  doi = {10.1109/MM.2020.2976067}
}

@article{minkenbergCopackagedDatacenterOptics2021,
  title = {Co-Packaged Datacenter Optics: Opportunities and Challenges},
  author = {Minkenberg, Cyriel and Krishnaswamy, Rajagopal and Zilkie, Aaron and Nelson, David},
  journal = {IET Optoelectronics},
  volume = {15},
  number = {2},
  pages = {77--91},
  year = {2021},
  doi = {10.1049/ote2.12020}
}

@article{rhimVerilogABehavioralModel2015,
  title = {{Verilog-A} Behavioral Model for Resonance-Modulated Silicon Micro-Ring Modulator},
  author = {Rhim, Jinsoo and Ban, Yoojin and Yu, Byung-Min and Lee, Jeong-Min and Choi, Woo-Young},
  journal = {Optics Express},
  volume = {23},
  number = {7},
  pages = {8762--8772},
  year = {2015},
  doi = {10.1364/OE.23.008762}
}

@article{zhangCompactPhotonicModel2024,
  title = {Compact Photonic Model Based on Coupled-Mode Theory for Nonlinear Interactions in Electronic-Photonic Co-Simulation},
  author = {Zhang, Siyuan and Fan, Xiaolong and Chen, Nuo and Qiu, C. and Wang, X. and Wang, Ken Xingze and Xu, J. and Tan, M.},
  journal = {Optics Express},
  volume = {32},
  number = {16},
  pages = {27599--27613},
  year = {2024},
  doi = {10.1364/OE.529044}
}

@inproceedings{zhangVerilogACompactModel2024,
  title = {A {Verilog-A} Compact Model for Four-Wave Mixing Supporting Electronic-Photonic Co-Simulation},
  author = {Zhang, Siyuan and Fan, Xiaolong and Chen, Nuo and Wang, Ken Xingze and Xu, J. and Tan, M.},
  booktitle = {Proc. Int. Symp. Electronics Design Automation (ISEDA)},
  pages = {16--19},
  year = {2024}
}

@article{ipCompensationDispersionNonlinear2008,
  title = {Compensation of Dispersion and Nonlinear Impairments Using Digital Backpropagation},
  author = {Ip, Ezra and Kahn, Joseph M.},
  journal = {Journal of Lightwave Technology},
  volume = {26},
  number = {20},
  pages = {3416--3425},
  year = {2008},
  doi = {10.1109/JLT.2008.927791}
}

@inproceedings{fougstedtTimedomainDigitalBack2017,
  title = {Time-Domain Digital Back Propagation: Algorithm and Finite-Precision Implementation Aspects},
  author = {Fougstedt, Christoffer and Mazur, Mikael and Svensson, Lars and Eliasson, Henrik and Karlsson, Magnus and Larsson-Edefors, Per},
  booktitle = {Proc. Optical Fiber Communication Conf. (OFC)},
  pages = {W1G.4},
  year = {2017},
  doi = {10.1364/OFC.2017.W1G.4}
}

@article{goldfarbChromaticDispersionCompensation2007,
  title = {Chromatic Dispersion Compensation Using Digital {IIR} Filtering With Coherent Detection},
  author = {Goldfarb, Gilad and Li, Guifang},
  journal = {IEEE Photonics Technology Letters},
  volume = {19},
  number = {13},
  pages = {969--971},
  year = {2007}
}

@article{madsenOpticalAllpassFilters1998,
  title = {Optical All-Pass Filters for Phase Response Design with Applications for Dispersion Compensation},
  author = {Madsen, C. K. and Lenz, G.},
  journal = {IEEE Photonics Technology Letters},
  volume = {10},
  number = {7},
  pages = {994--996},
  year = {1998}
}
\end{document}